\documentclass{jfm} 
\usepackage{amsmath,amssymb,latexsym, amscd}
\usepackage{exscale, graphics, epsfig, epstopdf, esint, color}
\usepackage{eucal}

\begin{document}

\shorttitle { On the dynamics of a free surface of an ideal fluid in a bounded domain.}
\shortauthor{S. A. Dyachenko}

\title { On the dynamics of a free surface of an ideal fluid in a bounded domain in the presence of surface tension.}

\author{Sergey A. Dyachenko
\corresp{\email{sdyachen@math.uiuc.edu}}
}

\affiliation{
Department of Mathematics, University of Illinois at Urbana-Champaign \\ Urbana, IL 61801 USA}

\maketitle
\begin{abstract}
We derive a set of equations in conformal variables that describe a potential flow of an ideal inviscid fluid 
with free surface in a bounded domain. 
This formulation is free of numerical instabilities present in the equations for the surface elevation and 
potential derived in~\cite{ZakharovEtAl96} with some of the restrictions on analyticity relieved. 
We illustrate with the results of a comparison of the numerical simulations with the exact solution, the 
Dirichlet ellipse (\cite{LonguettH1972}). In presence of surface tension, we demonstrate the oscillations of 
the free surface of a unit disc droplet about its equilibrium, the disc shape.
\end{abstract}

\section{Introduction}
From the theoretical point of view a droplet of ideal fluid is just as exciting and complicated object as an entire ocean.
It is quite common to study water waves on the surface of the ocean, we intend to demonstrate an efficient way to study water
waves on the free surface of ideal fliuid droplet subject to the force of surface tension.

The first classical results for the motion of ideal fluid over free surface have been obtained by ~\cite{Stokes1880}
and before that a class of exact solutions was found by ~\cite{Dir1860}. 
A detailed study of Dirichlet solutions including the ellipse, and the hyperbola was made in the work~\cite{LonguettH1972}.
In the second half of the twentieth century V. Zakharov discovered that the surface elevation and the velocity potential 
on the free surface are canonical Hamiltonian variables in the work~\cite{Zakharov1968}.
The conformal mapping approach for the Euler description of the full, non-stationary problem was first introduced by ~\cite{Tanveer1993}. 
In 1996, the works~\cite{ZakharovEtAl96} and~\cite{DyachenkoEtAl96} established a formulation for non-stationary 
water wave based on a conformal mapping of the fluid domain to lower complex plane. This formulation has been proven quite successful 
for numerical simulations using pseudospectral method, although the equations have been observed to suffer from 
numerical instability due to truncation of the Fourier series. The work~\cite{Dyachenko2001} offered a reformulation that is 
free from numerical instabilities and contains only polynomial nonlinear terms in the equation. A sequence of works has 
followed that was employing this formulation see ~\cite{ZakharovEtAl2002}. In 2009 the formulation has been adapted to handle 
a bubble of air encircled by the fluid in the work~\cite{TuritsEtAl2009}.

The Dirichlet ellipse is one of the exact solutions where fluid volume is bounded. It is the natural candidate to test the validity of 
the newly derived equations. We simulate the motion of the Dirichlet ellipse and have observed excellent agreement with the analytical
solution. 

\section{Formulation of the problem}   
We study two--dimensional incompressible fluid that fills a bounded domain $\mathcal{D}$. 
The boundary of the fluid domain, $\partial \mathcal{D}(t)$, is a free surface given in the implicit form, $F(x,y,t) = 0$. 
The fluid flow is potential and the velocity field is given by ${\bf v}(x,y,t) = \nabla \varphi(x,y,t)$ and 
\begin{align}
\nabla^2 \varphi = 0. \nonumber
\end{align}
The non-stationary Bernoulli equation governs the evolution of the velocity potential, in particular at $\partial \mathcal{D}$:
\begin{align}
\left.\left(\varphi_t + \frac{1}{2}\left( \nabla \varphi \right)^2 + p\right)\right|_{F(x,y,t)=0} = 0,\label{eqn8}
\end{align}
where $p = p(x,y,t)$ is the pressure. We neglect the atmospheric pressure at the free surface, and have 
$p = \sigma \kappa$, where $\sigma$ is the surface tension coefficent, and $\kappa$ is the local curvature. 

In general, a fluid particle on the free surface moves in both the tangential and the normal direction, however it is only the motion in the 
normal direction that changes the shape of the fluid boundary. This motion is captured by the kinematic boundary condition:
\begin{align}
\dfrac{\partial F}{\partial t} + \left.\left(\nabla \varphi, {\bf n} \right)\right|_{F(x,y,t)} = 0, \label{eqn0}
\end{align}
where ${\bf n} = {\bf n}(x,y,t)$ is the unit normal to $\partial \mathcal{D}$ at the point $(x,y)$. 

The total energy, $\mathcal{H}$, associated with the fluid flow is given by the sum of the kinetic energy, $T$, and the potential 
energy, $\mathcal{U}$:
\begin{align}
T = \frac{1}{2}\iint\limits_{\mathcal{D}} \left( \nabla \varphi \right)^2 \,dxdy, \quad 
\mathcal{U} = \sigma \int\limits_{\partial D} dl, \nonumber
\end{align}
where $\sigma$ is the coefficient of surface tension, and $dl$ is the elementary arclength along $\partial \mathcal{D}$.

\begin{figure}
\includegraphics[width=0.98\textwidth]{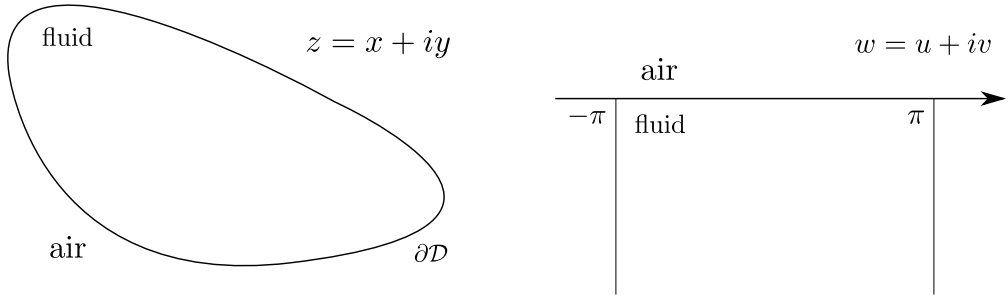}
\caption{ Illustration of the conformal map: (left) the fluid domain $z=x+iy$ is enclosed by the 
free surface $\partial \mathcal{D}$, and (right) the conformal domain $w = u+iv$ is the periodic strip in the lower 
complex half--plane.}
\label{Fig0}
\end{figure}
\section{Conformal Variables.}

Let $z(w) = x(w) + iy(w)$ be a conformal map to the fluid domain $(x,y)\in \mathcal{D}$ from a periodic strip 
${w = u+iv, -\pi \leq k_0 u < \pi, v \leq 0}$ and $k_0$ is the base wavenumber for the parameterization of the circle.  
The illustration of the mapping is given in the Figure~\ref{Fig0}. The free surface, $\partial \mathcal{D}$ is mapped 
from $v = 0$ and satisfies:
\begin{align}
F(x(u,t), y(u,t), t) = 0. \label{eqn1}
\end{align}
The kinematic condition can be revealed from the observation of the time--derivative of the implicit function $F$:
\begin{align}
\frac{dF}{dt} = F_t + F_x x_t + F_y y_t = 0,\label{eqn2}
\end{align}
where subscript denotes a partial derivative. After a change of coordinates from the $(x,y)$--plane to the $(u,v)$--plane, 
and applying the chain rule it becomes:
\begin{align}
F_t +  \frac{x_t x_u + y_t y_u}{|z_u|^2}F_u + \frac{x_ty_u - y_t x_u}{|z_u|^2} F_v = 0, \label{eqn3}
\end{align}
and the coefficient at $F_v$ gives the rate of change of free surface $\partial \mathcal{D}$ in the vertical direction in $w$--plane,
which translates to the normal direction in the $(x,y)$--plane.
In conformal variables the equation~\eqref{eqn0} can be written by noting that the unit normal ${\bf n}$ is given by:
\begin{align}
{\bf n} = \frac{1}{|z_u|^2}\left(-y_u, x_u \right)^T, \nonumber
\end{align}
where we exploit the fact that $z(w)$ is subject to the Cauchy--Riemann (CR) relations. A change of variables in the~\eqref{eqn0} thus 
reveals that:
\begin{align}
F_t + \frac{\psi_v}{|z_u|^2} F_v = 0,\label{eqn35}
\end{align}
where we have introduced a restriction of the velocity potential to the free surface:
\begin{align}
\psi(u,t) = \left.\varphi(x,y,t)\right|_{F(x,y,t) = 0}. \nonumber
\end{align}
By matching the coefficient at the partial derivative $F_v$ in the equations~\eqref{eqn3} and~\eqref{eqn35} we discover the kinematic condition in conformal domain:
\begin{align}
\frac{x_t y_u - y_t x_u}{|z_u|^2} = \frac{\psi_v}{|z_u|^2}, \label{eqn4}
\end{align}
it ensures that the line $v = 0$ maps to the free surface for all time. To have a derivative with respect to $v$ is a nuisance 
that can be alleviated by making use of the stream function, $\theta$, and using CR relations together with the result of Titchmarsh theorem 
to find that $\psi_v = -\hat H \psi_u$, where $\hat H$ denotes the Hilbert transform:
\begin{align}
\hat H f(u) = v.p. \int\limits_{-\infty}^{\infty} \frac{f(u')\, du'}{u' - u}, \nonumber
\end{align}
where $v.p.$ denotes a Cauchy principal value integral.

The Bernoulli equation~\eqref{eqn8} determines the evolution of the velocity potential at the free surface. It is formulated 
in conformal variables by means of elementary calculus. The force of surface tension is proportional to the local curvature 
of $\partial \mathcal{D}$:
\begin{align}
p &= -\sigma \frac{x_u y_{uu} - y_u x_{uu}}{|z_u|^3}. \label{eqn9} 
\end{align}
Under the conformal change of variables the surface potential, $\psi(u,t)$ becomes a composite function:
\begin{align}
\psi(u,t) = \varphi(x(u,t),y(u,t),t), \nonumber 
\end{align}
whose time--derivative together with the equations~\eqref{eqn8},~\eqref{eqn4} and~\eqref{eqn9} reveal:
\begin{align}
\psi_t = -\frac{\psi_u^2 - \left(\hat H \psi_u \right)^2}{2|z_u|^2} + \psi_u \hat H \left[\frac{\hat H \psi_u}{|z_u|^2}\right] 
- p, \label{eqn10}
\end{align}
the dynamic boundary condition in the conformal domain.
\section{The complex equations.}
In order to reveal the analytic structure of the problem at hand, it is convenient to introduce the 
complex potential, $\Phi = \psi + i\theta$. By the Titchmarsh theorem the complex potential can also 
be written in the form:
\begin{align}
\Phi = \psi + i \hat H \psi = 2\hat P \psi,
\end{align}
where $\hat P$ is the projection operator. The both complex functions $\Phi$ and $z$ are analytic 
in the periodic strip in the lower complex half--plane. The free surface, $\partial D$ is a closed curve in the 
$(x,y)$--plane, and hence the conformal map $z(u + iv)$ must be a periodic function of the variable $u$. Therefore 
$z(w)$ (as well as $\Phi(w)$) can be expanded in Fourier series, and the analyticity in the periodic strip 
requires that only the nonpositive Fourier coefficients are nonzero:
\begin{align}
z(w) = \hat z_0 + \sum\limits_{k_0 \mathbb{N}} \hat z_k e^{-ikw},
\end{align}
where $\hat z_k$ denote the Fourier coefficients of $z$. As evident from this expansion as $w\to -i\infty$ the derivative 
of the conformal map $z_w \to 0$ and $z(w\to-i\infty) = z_0$. In other words, there exists an analytic function 
$\rho(w)$, that satisfies:
\begin{align}
\rho z_w = e^{-ik_0w}
\end{align}
at every point in the strip, and $\rho(-i\infty) = (-ik_0\hat z_1)^{-1} \neq 0$. We will also introduce a zero--mean, analytic 
function, $\nu = i \rho \Phi_u$. When the kinematic condition is written in the complex form and multiplied by $|\rho|^2$, 
the choice of $\rho$ and $\nu$ becomes transparent: 
\begin{align}
&z_t \bar z_u \rho \bar \rho - \bar z_t z_u \rho \bar \rho = \bar \Phi_u \rho \bar \rho - \Phi_u \rho \bar \rho,\nonumber\\
&\rho z_t e^{ik_0 w}    - \bar \rho \bar z_t e^{-ik_0 w} = i\left( \rho \bar \nu +  \bar \rho \nu \right), \nonumber
\end{align}
where $z_t\left(w\to-\infty \right) \to 0$ and the left--hand side is the difference of two complex analytic functions.
We apply the projection operator $\hat P$ to have
\begin{align}
\rho z_t e^{ik_0 u} = i\hat P \left[ \rho \bar \nu +  \bar \rho \nu \right].\nonumber
\end{align}
After elementary calculation, we conclude that the analytic function $\rho$ satisfies the pseudo--differential equation:
\begin{align}
\rho_t = i \left( U \rho_u - U_u \rho  \right) - k_0\rho U, \label{eqn7}
\end{align}
where $U$ is given by:
\begin{align}
U = \hat P \left( \rho \bar \nu +  \bar \rho \nu \right) \nonumber
\end{align}
is the complex transport velocity.
The equation for the complex potential is found by applying $2\hat P$ to the equation~\eqref{eqn10}:
\begin{align}
\Phi_t = i U \Phi_u - B,\nonumber
\end{align}
where $B$ is given by:
\begin{align}
B &= \hat P \left[ \frac{|\Phi_u|^2}{|z_u|^2} -2\sigma \frac{x_uy_{uu} - y_u x_{uu}}{|z_u|^3} \right], \nonumber \\  
B &= \hat P \left[ |\nu|^2 + 2\sigma k_0 |\zeta|^2 + 2i\sigma \left( \zeta \bar \zeta_u - \bar \zeta \zeta_u \right)  \right], \nonumber
\end{align}
where $\zeta^2 = \rho $ has been derived in the reference~\cite{ZakharovEtAl96}. 
We omit the trivial details of calculation and skip to the result, which is the equation satisfied by $\nu$:
\begin{align}
\nu_t = i \left( U \nu_u - B_u \rho \right) - k_0 U\nu. \label{eqn11}
\end{align}
The equation~\eqref{eqn11} is analogous to the one originally discovered in the reference~\cite{Dyachenko2001} 
for the infinite fluid domain. Together with the equation~\eqref{eqn7}, the equation~\eqref{eqn11} forms a closed 
system suitable for numerical simulation. 

\section{The choice of the reference frame.}

The equations~\eqref{eqn7},~\eqref{eqn11} are formulated for the derivative of conformal map, and henceforth 
they describe only the motion of the fluid relative to the point of the fluid, $\hat z_0$. The motion of this 
point, namely $\hat z_0(t) = z(-i\infty)$ is not captured in the equations~\eqref{eqn7},~\eqref{eqn11} and has to 
be recovered from elsewhere. This missing puzzle piece comes from the momentum conservation. Since the center of 
mass of the fluid is an inertial reference frame, it is convenient to choose it to be at rest at the origin. 
The location of the center of mass is given by:
\begin{align}
\iint\limits_{\mathcal{D}} z\,dxdy = \iint z |z_u|^2 \,du dv = 0. \label{eqn12}
\end{align}
With the total mass of the fluid being a constant of motion given by:
\begin{align}
m = \iint\limits_{\mathcal{D}}\,dxdy, \nonumber
\end{align}
the equation~\eqref{eqn12} can be written as:
\begin{align}
m \hat z_0 = -\iint \left( z - \hat z_0\right) | z_u|^2 \,du dv, \label{eqn13}
\end{align}
and is used to recover the zero Fourier mode of the conformal map, quite similar to the zero mean condition 
that is often imposed in the problem with infinite fluid domain. Together with the system~\eqref{eqn7},~\eqref{eqn11}, the 
equation~\eqref{eqn13} fully describes the motion of the boundary of the fluid, $\partial \mathcal{D}$.

\begin{figure}
\includegraphics[width=0.49\textwidth]{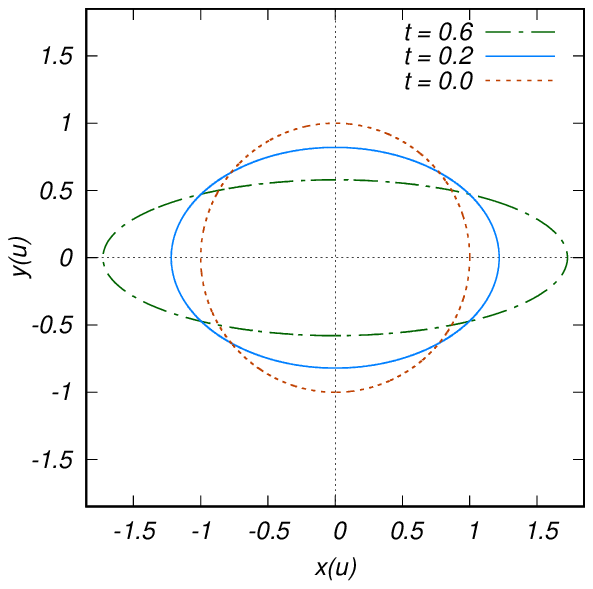}
\includegraphics[width=0.49\textwidth]{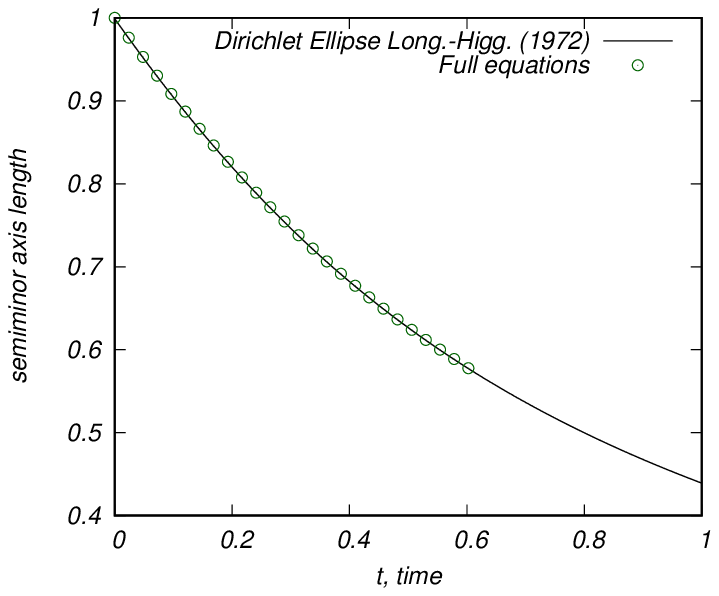}
\caption{The shape of the fluid boundary at time $t = 0$ (dashed), $t = 0.2$ (solid) and $t = 0.6$ (dash--dotted) as found from numerical 
simulations of the~\eqref{eqn7} and ~\eqref{eqn11} with initial conditions in accordance with Dirichlet ellipse~\eqref{SV_init}.}
\label{Fig2}
\end{figure}

The equations~\eqref{eqn7},~\eqref{eqn11} for the analytic functions $\rho$, $\nu$ are equivalent to the equations for 
$R = \frac{1}{z_u}$ and $V = i\Phi_u R$:
\begin{align}
R_t = i \left( UR_u - U_u R\right), \quad V_t = i\left(U V_u - B_u R\right), \nonumber
\end{align}
with
\begin{align}
U &= \hat P\left[ V\bar R + \bar V R \right], \nonumber \\
B &= \hat P\left[ |V^2| + 2i\sigma \left( Q \bar Q_u - \bar Q Q_u \right)\right] \nonumber
\end{align}
where $Q^2 = R$.

There is, however, an important caveat, namely the function $R$ is no longer analytic at $w\to -i\infty$, 
but instead it contains a term in its Fourier series expansion with a positive wavenumber, $k_0$:
\begin{align}
R(w) = e^{ik_0 w}\left(\hat \rho_0 + \sum\limits_{k_0\mathbb{N}} \hat \rho_k e^{-ikw} \right) = \rho e^{ik_0w}. \nonumber
\end{align}


\section{Numerical Experiments}
In the remainder of the text we demonstrate the simulations of the equations~\eqref{eqn7} and~\eqref{eqn11} and compare the results 
with available exact solution. The simulations are performed on a uniform grid in the $u$--variable using Runge-Kutta fourth 
order timestepping scheme. The spatial derivative, $\partial_u$, and projection operator, $\hat P$ are applied as Fourier 
multipliers to the coefficients of Fourier series of the respective functions.

In the first simulation we compare the solutions of~\eqref{eqn7},~\eqref{eqn11} with Dirichlet ellipse. In the Dirichlet 
ellipse the complex potential, $\Phi$ is a quadratic function of the conformal map $z$:
\begin{align}
\Phi = \frac{1}{2} Az^2 + \int f dt, \nonumber
\end{align}
where $A = A(t)$ and $\int f\,dt$ is the Bernoulli constant. The surface shape, $\partial \mathcal{D}$ is 
given by:
\begin{align}
\frac{x^2}{a^2} + \frac{y^2}{b^2} = 1, \nonumber
\end{align}
where $a = a(t)$ and $b = b(t)$ are determined from $A$ as follows:
\begin{align}
a^2 = \frac{1}{b^2} = \frac{A^2}{1-\sqrt{1-A^2}}, \nonumber
\end{align}
and $A$ satisfies an ordinary differential equation (ODE):
\begin{align}
\dfrac{dA}{dt} = A^2 \sqrt{1-A^4}. \label{DirichletODE}
\end{align}
This ODE is solved numerically and is compared to the solution of~\eqref{eqn7},~\eqref{eqn11} and 
agreement is demonstrated in Figure~\ref{Fig2} (left panel). We measure the size of the semiminor axis, $b(t)$ as obtained
from both simulations and plot the result in Figure~\ref{Fig2} (right panel). The initial data for the simulation is given by
\begin{align}
\rho(t = 0) = 1, \quad \nu(t = 0) = -ie^{-iu}, \label{SV_init}
\end{align}
then 
\begin{align}
z(t = 0) = e^{-iw}, \quad \Phi(t = 0) = \frac{1}{2}z^2. \nonumber
\end{align}

\begin{figure}
\includegraphics[width=0.49\textwidth]{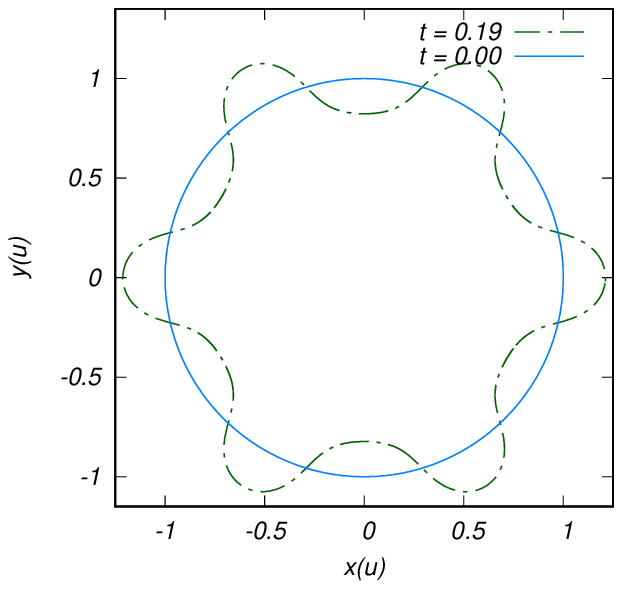}
\includegraphics[width=0.49\textwidth]{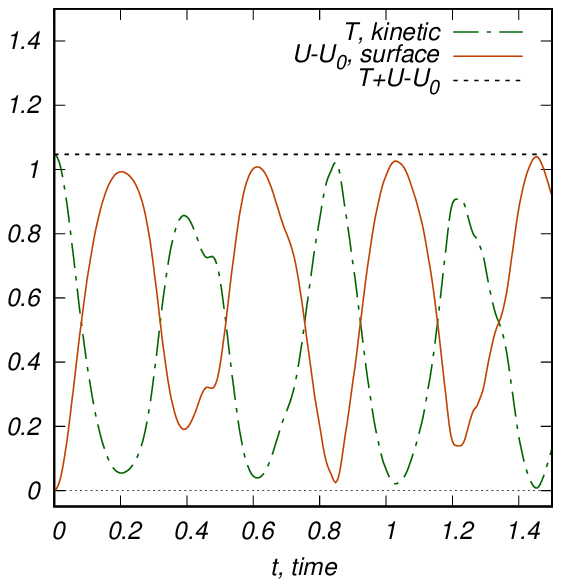}
\caption{Numerical simulation of the motion of droplet with surface tension coefficient $\sigma = 0.5$: 
the initial shape of the surface (solid line) and the shape of the surface at the maximal value of surface energy, at approximately $t = 0.19$ (dash--dotted line).}
\label{Fig3}
\end{figure}

In the second simulation, we solve the equations~\eqref{eqn7} and~\eqref{eqn11} in presence of surface 
tension with $\sigma = \frac{1}{2}$ and take the initial data for complex potential to be:
\begin{align}
z(t = 0) = e^{-iw},\quad \Phi(t = 0) = -\frac{2}{5}e^{-5iw}, \nonumber
\end{align}
so the initial surface shape is a unit disk. In the course of the experiment the surface shape exhibits 
oscillatory motion quite similar, but not exactly that of a standing wave. We measure the accuracy of the simulations by verifying 
that the fluid mass, $m$, and the Hamiltonian, $\mathcal{H}$ are conserved to $14$ digits of precision. In
addition, we track the Fourier spectrum of the functions $\rho$ and $\nu$ to ensure that it is resolved. In the 
Figure~\ref{Fig3} we illustrate the shape of the surface at initial time $t = 0$ and at the time of the first 
local minimum of the kinetic energy at approximately $t = 0.19$.

\section{Conclusion}
A conformal mapping formulation that has been discovered in~\cite{Dyachenko2001} 
is extended to a problem of finite, simply connected fluid domain. The resulting equations are in agreement with 
the aforementioned work with some restrictions on analyticity lifted.
We demonstrate simulations without surface tension, the Dirichlet ellipse, and with surface 
tension. It is worthwhile to point out that it is typically not recommended to perform simulations 
on uniform numerical grid in the $u$--variable, instead there are techniques to speed--up Fourier 
series convergence, for details see~\cite{LushEtAl2017}.

The presented work is a precursor to a further study of the droplet splitting problem using conformal approach. 
A Dirichlet hyperbola is a particularly promising candidate for the droplet splitting via a finite-time 
singularity formation and is the subject of an ongoing study. 

Another subject of the ongoing research is the motion of the perturbed boundary of the unit disc in presence of 
constant vorticity. The mathematical formulation of this problem is tractable in infinite fluid, see 
e.g.~\cite{ConstantinEtAL2004} for travelling wave solutions, and~\cite{DyachenkoEtAl2018} for the 
full time--dependent formulation in the conformal variables. It is percieved that the generalization 
to the droplet will be tractable as well, and furthermore the constant vorticity in a droplet carries 
more physical meaning than in an infinite depth domain, where it implies unbounded fluid velocities.

\section{Acknowledgements}
The author would like to express gratitude to Alexander Dyachenko, Vera Mikyoung Hur and Fabio Pusateri for fruitful 
discussions. The author thanks the creators and maintainers of the FFTW library~\cite{FFTW}
and the entire GNU project. This work was supported by NSF grant DMS-$1716822$.

\bibliographystyle{jfm}
\bibliography{droplet}

\end{document}